# An Efficient Privacy-Preserving Algorithm based on Randomized Response in IoT-based Smart Grid

Hui Cao
*Computer School*
*Wuhan University*
Wuhan, China
2013102110071@whu.edu.cn

Shubo Liu
*Computer School*
*Wuhan University*
Wuhan, China
liu.shubo@163.com

Zhitao Guan
*School of Control and Computer Engineering*
*North China Electric Power University*
Beijing, China
guan@ncepu.edu.cn

Longfei Wu
*Department of Mathematics & Computer Science*
*Fayetteville State University*
Fayetteville, NC, USA
lwu@uncfsu.edu

Haonan Deng
*Computer School*
*Wuhan University*
Wuhan, China
dhnzzz@126.com

Xiaojiang Du
*Department of Computer and Information Sciences*
*Temple University*
Philadelphia, PA, USA
xjdu@temple.edu

*Abstract*—Among existing privacy-preserving approaches, Differential Privacy (DP) is a powerful tool that can provide privacy-preserving noisy query answers over statistical databases and has been widely adopted in many practical fields. In particular, as a privacy machine of DP, Randomized Aggregable Privacy-Preserving Ordinal Response (RAPPOR) enables strong privacy, efficient, and high-utility guarantees for each client string in data crowdsourcing. However, as for Internet of Things(IoT), such as smart gird, data are often processed in batches. Therefore, developing a new random response algorithm that can support batch-processing tend to make it more efficient and suitable for IoT applications than existing random response algorithms. In this paper, we propose a new randomized response algorithm that can achieve differential-privacy and utility guarantees for consumer's behaviors, and process a batch of data at each time. Firstly, by applying sparse coding in this algorithm, a behavior signature dictionary is created from the aggregated energy consumption data in fog. Then, we add noise into the behavior signature dictionary by classical randomized response techniques and achieve the differential privacy after data re-aggregation. Through the security analysis with the principle of differential privacy and experimental results verification, we find that our Algorithm can preserve consumer's privacy without comprising utility.

## I. INTRODUCTION

Differential privacy [1-6] constitutes a strong standard for privacy guarantees in statistical databases. Due to its property of rigid and provable privacy guarantee, differentially private database mechanisms can make confidential data widely available for accurate data analysis and data statistics [3]. Although the traditional Laplace mechanism is applicable to numerical query results, in many practical applications, query results are logical objects [7-8]. A recent privacy-preserving crowdsourcing mechanism, Randomized Aggregable Privacy-Preserving Ordinal Response (RAPPOR) [9], is proposed in 2014, which has substantially improved the statistical crowdsourcing technology. Since then, RAPPOR has been adopted and enhanced by many researchers to generate specialized Randomized Response algorithms that can be used in specific fields, such as machine learning [10-15].

RAPPOR may be used to solve the problem of privacy protection in smart grid which shares some similarity with the crowdsourcing environment. As the next-generation power system, smart grid is a special Internet of things system that that envisioned to provide efficient and reliable energy delivery [16]. To realize fine-grained optimization and real-time management for energy supply, numerous smart meters are installed at the consumers' home to gather the energy consumption data at high frequency and upload the data to the control center [17]. However, these detailed data usually include privacy-sensitive information with regard to the consumer's application consumption patterns, which raises the consumer's behavior privacy concerns [18].

Actually, there are numerous works contributing on the privacy preservation in smart grid. Traditional schemes mainly contain the following three aspects. Firstly, Homomorphic encryption [19-21] is a common solution to preserve consumer's privacy. However, it introduces a heavy computational overhead and always needs a third party for random secret sharing distribution [22]. Secondly, battery-based load hiding (BLH) method is to hide the actual electricity consumption data of appliance by charging and discharging the battery [23-27], but the environmental effects and cost caused by chargeable battery can't be ignored.

DP has been applied in IoT to protect customers' privacy. Existing differential privacy schemes in smart grid mainly focus on the individual energy consumption data [28-32] and adding noise into the energy consumption data is the common solution to provide differential privacy. Nevertheless, the noisy data may affect the data-utility and have a bad impact on several services such as electricity billing in smart grid.

RAPPOR provides strong privacy, efficient, and high-utility guarantees for each client string in data crowdsourcing. It is an intuitive idea to apply RAPPOR for differential privacy in IoT context. However, in this case, both Basic RAPPOR or One time RAPPOR need to use Bloom filter that can hash a string or very few bits each one time. In consideration of massive batches of data, there are difficult to address, potentially causing surveyors to adopt the less effective privacy protections [11].

In the IoT and fog computing environment, the terminal devices often need to process data by batch. For example, smart meter collects energy consumption data and send it every 15 minutes. Although RAPPOR is a good machine of differential privacy, it is hard to process data in such volume.

In this paper, we propose an improved randomized response algorithm that could protect consumer's privacy from behavior signatures and our contributions are summarized as follows:

(1) The proposed algorithm is applicable for privacy protection of IoT. Differing from traditional differential private approaches, we add randomized response noise into the behavior signatures matrix to achieve an acceptable utility-privacy tradeoff.

(2) The key of the design is that we propose a behavior signature modeling method based on sparse coding instead of boom filter. After some lightweight trainings using the energy consumption data, the dictionary will be associated with the behavior characteristics of the electric appliances. The dictionary learning is sparse.

(3) At last, the performance of our scheme is compared with the state of the arts. The experimental results show that our algorithm can achieve a good balance between data-utility and privacy, hence is better applicable to IoT applications.

The rest of this paper is organized as follows. Section II introduces the related work and background. In section III, our algorithm is presented in detail. In section IV, security and utility analysis are given. In section V, the performance of our scheme is evaluated. In section VI, the paper is concluded.

## II. RELATED WORK

### A. Differential privacy

Dwork [1] has proposed the notion of differential privacy for privacy-preserving in statistical datasets. In addition, he has also proposed how to realize differential privacy by noise addition [33]. McSherry [34] studied the property of the parallel composition and stable transformation in differential privacy. Kifer [8] analyzed the privacy-utility tradeoff and provided the metrics for data-utility. Differential privacy in smart grid has been discussed in [28-32], and the fault-tolerance problem during data aggregation has been analyzed in [31] [35]. Barbosa [36] firstly applied differential privacy against NILM (Non-Intrusive Load Monitoring) to protect the data collected by smart meters.

1) Definition of differential privacy

$M$ is a privacy mechanism. For any datasets $D_1$ and $D_2$, differing from at most one element, M satisfies $\varepsilon$-differential privacy if the two datasets satisfy the following condition:

$$P_r\{M(D_1) \subseteq S\} \le e^{\varepsilon} \times P_r\{M(D_2) \subseteq S\}$$

$S$ represents the range of $M(D_i)$. The smaller the value of $\varepsilon$ is, the higher the degree of privacy preservation can be achieved.

2) Property1: Parallel Composition [34]

$M_1, M_2...M_n$ are different privacy mechanisms with the privacy budgeting parameters $\varepsilon_1, \varepsilon_2...\varepsilon_n$. Then, the combined algorithm $M(M_1(D_1), M_2(D_2)...M_n(D_n))$ provides $(\max \varepsilon_i)$-differential privacy for the disjoint datasets $D_1, D_2...D_n$.

3) Property2: Stable Transformations [34]

For any two databases E and F, we say T provides c-stable if it meets the following condition.

$$|T(E) \oplus T(F)| \le c \times |E \oplus F| \qquad (1)$$

If the privacy preserving mechanism M provides $\varepsilon$-differential privacy and $T$ is a c-stable transformation, the combination of $M \times T$ provides $(\varepsilon \times c)$-$\varepsilon \times$ c differential privacy.

### B. Randomized responses

Randomized Response [43] is a survey technique developed in the 1960s for collecting statistics on sensitive topics where survey respondents wish to retain confidentiality.

Erlingsson[9] proposed RAPPOR as a differential privacy machine, and answered how to recover distributional parameters from the responses. RAPPOR becomes a widely used differential privacy approach. The consideration of several bits jointly, had to wait for a subsequent operation, Fanti proposed an improved Algorithm [10] to deal with several bits jointly. Staal A. Vinterbo extended RAPPOR to as n-dimensional[11]

RAPPOR performs the following steps: Firstly, hash client's value onto the Bloom filter.

Then let

$$x_i' = \begin{cases} 1, & withprobability \ \frac{1}{2}f \\ 0, & withprobability \ \frac{1}{2}f \\ x_i, & withprobability \ 1-f \end{cases}$$

$x_i$ represents the original energy consumption data.

$x_i'$ represents the data after being obfuscated by RAPPOR.

### C. Sparse Coding

Sparse coding is a kind of method for learning sets of over-complete bases to represent data efficiently. It has multiple variances such as hierarchical sparse coding [37], non-negative sparse coding [38] and has been applied in many fields like NILM, audio classification [39] and source separation [40]. In the field of smart grid, Kolter [37] extended the sparse coding to improve the accuracy of a large-scale energy disaggregation task and Elhamifar's [40] designed a method based on sparse coding to extract signature consumption pattern. Jenatton [41] proposed an optimal method for the regularization of hierarchical sparse coding. Singh [42] proposed a greedy deep sparse coding for energy disaggregation.

The aim of sparse coding is to find dictionary $B_i \in R^{r \times n}$ such that the input vector $x_i$ can be represented as a linear combination of $n$ basis functions as follows.

$$x_i \approx B_i A_i$$

Here, $B_i$ denotes the dictionary and the number of basis function $n$ is larger than the dimensionality of data $x_i$. The activation matrix $A_i \in R^{n \times 1}$ is sparse, which means the sets of activations contain mostly zero entries. The sparse coding cost function is defined as follows.

$$\min_{A_i \geq 0, B_i \geq 0} \| x - B_i A_i \|_F^2 + \lambda \sum_{p,q} (A_i)_{pq}$$

$\| Y \|_F$ represents the Frobenius norm of $Y$ and the value of $F$ is always set to two in sparse coding. $C$ represents a constant. $B_i^j$ denotes the $j$th column of matrix $B_i$. For a dataset $\hat{X}$ composed of $v$ elements, the sparse coding can be shown as follows. $\lambda$ denotes the weight of sparsity.

$$A_{1:v} = \arg\min_{A_{1:v} \geq 0} \left\| \hat{X} - [B_1 ... B_v] \begin{bmatrix} A_1 \\ ... \\ A_v \end{bmatrix} \right\|_F^2 + \lambda \sum_{i,p,q} (A_i)_{p,q}$$

## III. FRAMEWORK AND ALGORITHM

### A. Design Goal

In general, as shown in Figure 1, the objective of the attacker is to infer household activity from the power consumption data.

Specifically, the attack model is
(1) With the total energy consumption of all consumers in a region at a certain time, inferring the consumption of a certain consumer.
(2) With the total energy consumption of a consumer over a period of time, inferring the behavior of that consumer.

Inherited from Barbosa's [36] design goals, our schemes focus on the following aspects:
(1) Enabling the calculation of the total consumption of a consumer over a period of time (*e.g.*, monthly billing);
(2) Enabling the calculation of the total consumption of all consumers in a region at a certain instant of time;
(3) Avoiding the measurement of the instantaneous consumption of an individual consumer at a certain instant of time.

Besides, we also propose two new design goals.
(4) Providing $\varepsilon$-differential privacy without generating a serious impact on the data-utility.
(5) The obfuscated data should not be far from the original data in case that the adversary removes the noise to get the real data.

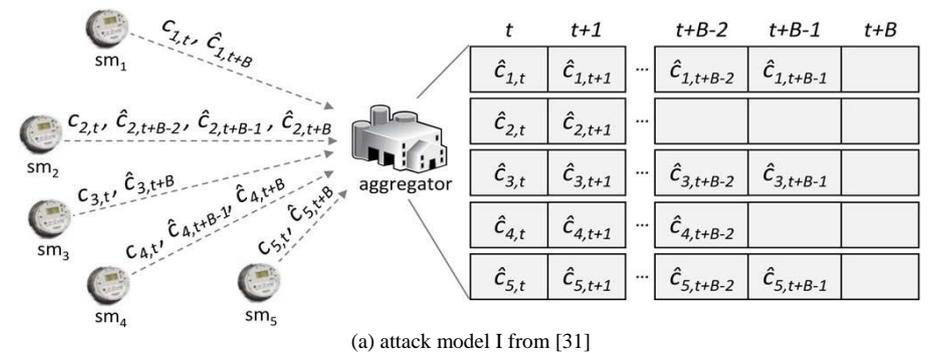

(a) attack model I from [31]

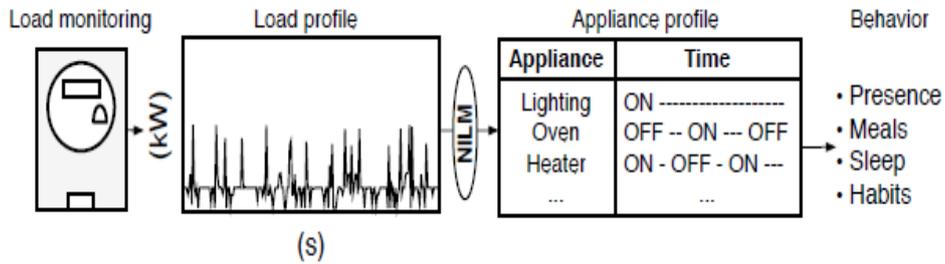

(b) attack model II from [44]

Figure.1. As shown in (a), for some consumers in a region at a certain instant of time, the adversary may learn a certain consumer's consumption data and then infer the consumer's behaviors and habits.. As shown in (b), for a consumer over a period of time, the adversary may learn the switch states of each appliance by NILM and then infer the consumer's behaviors and habits.

## B. System Framework

As shown in figure.2, the algorithm takes in the real energy consumption data, and performs the following steps:

(1) Train the dictionaries (basis functions)

This step can be complete by a trusted fog using historical data as the training set. If the electric utility data is trusted by customers, the dictionaries can be trained from electric utility data that are sent to the fog. Although the cost of training may be time-consuming, this work does not need to be carried out repeatedly, and it can be completed at the more powerful fog layer.

(2) Sparse coding

This step can be completed by the trusted fog regularly (e.g., 15 minutes), along with the data collected from the smart meter.

(3) Randomized response

This step can be completed by the trusted fog.

(4) Data re-aggregation.

This step can be completed by the trusted fog.

(5) Report. Send the generated report S to the server.

This step can be completed by the trusted fog. An end-to-end connection between the smart meter and the utility that could send noisy data.

## C. Sparse coding

The energy consumption data $Y_{1\times t}$ can be disaggregated by sparse coding as follows:

$$Y'_{t\times 1} = B_{t\times n}A_{n\times 1} \approx Y_{t\times 1} \quad (2)$$

Here, $t$ denotes the amount of time slots, which is one deal in whole date set. $n$ is the number of basic functions $b$, which is larger than $t$. The dictionary $B_{t\times n} = [b_1...b_n]$ is related to the energy consumption dynamics corresponding to the different operation modes of a device. $A_{n\times 1}$ is the activation matrix.

Dictionaries are the critical factor to sparse coding.

In IoT, efficient dictionary should be associated with the energy consumption data, Switch state of the appliance and the behavior signature. Energy consumption data is collected from smart meters, which is a common IoT application considered in many schemes such as [17] [18]. The switch state of an appliance is the appliance consumption patterns (ON/OFF) of a particular consumer [38]. The privacy-preserving requirement is that the application consumption patterns should not reflect a particular consumer's behavior at home, such as presence or absence, sleep-wake-cycles and other personal behaviors.

To achieve the desired goal using sparse coding, two important conditions must be satisfied:

- The combining error must small.

$$B_{t\times n}A_{n\times 1} - Y_{t\times 1} \to 0$$

- The Activations matrix $A_{n\times 1}$ is sparse

$$A_{n\times 1} = \{a_1, a_2, ..., a_i\}$$

Element $a_i$ should be as close to 0 as possible.

It's obvious that the random initial dictionary $B$ is far away from this goal, hence a continuous optimization process is required.

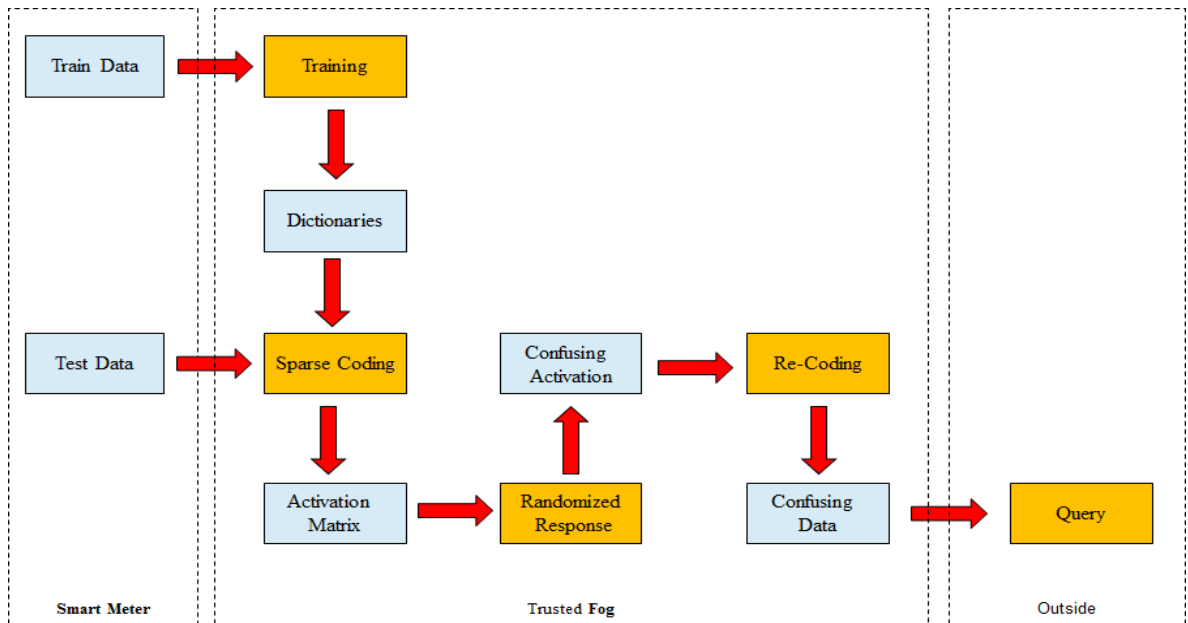

Figure.2. The process of our algorithm, training data, testing data and obfuscated data are represented by pale blue. The data processing like supervised learning, modified sparse coding, noise addition and data re-aggregation is represented by orange. Basic functions (dictionaries), aggregated activations and obfuscated activations belong to the intermediate data. Firstly, we train the basic functions based on supervised learning. Then, we predict the aggregated activations based on modified sparse coding from the testing data. At last, noise is added into activation and we get the obfuscated data by data re-aggregation.

## D. Training dictionary

The process of training the dictionary is described as follows:

Firstly, define a suitable objective function with the dictionary $B$ and activation $A$. Then, the objective function iterates, which solidifies and improves the dictionary $B$ and activation $A$ alternatively (solidifies the activation $A$ and improves dictionary $B$, then solidifies the dictionary $B$ and improve activation $A$). This process is repeated until convergence, in which dictionary $B$ is appropriate.

The objective function is defined as follows.

$$\| Y_{t \times 1} - B_{t \times n} A_{n \times 1} \|^2_2 + \lambda f(A_{n \times 1}) \tag{3}$$

Here, $\| Y \|_F$ represents the Frobenius norm of Y and the value of $F$ is always set to 2 in sparse coding. $C$ represents a constant. $B_i^j$ denotes the *jth* column of matrix $B_i$. $\hat{X}$ is a dataset composed of $v$ elements. $\lambda$ denotes the weight of sparsity.

Here, $\| Y_{t \times 1} - B_{t \times n} A_{n \times 1} \|^2_2$ represents the cost of data-reconstruction.

$f(A_{n \times 1})$ is the restrictive condition of activation matrix $A_{n \times 1}$, where $f(A_{n \times 1})$ is a sparsity cost function which penalize $a_i$ for being far from zero. A typical choice is:

$f(A_{n \times 1}) = \|A_{n \times 1}\|_1$.

$\| A_{n \times t} \|_1$ captures the sparsity condition

The objective function based on sparse coding for the aggregated data sampling from $v$ devices can be defined to be

$$\min \left\| Y_{t \times 1} - [b_1 ... b_n] \begin{bmatrix} a_1 \\ ... \\ a_t \end{bmatrix} \right\|^2_2 + \lambda \sum_{i=1}^{t} \| A \|_1 \tag{4}$$

Given an energy consumption vector $Y_{t \times 1}$, to generate an acceptable dictionary, we train the dictionary as follows:

Firstly, we give an initial dictionary $\hat{B}_0$. According to the energy consumption profile of each appliance' state, we get vector set $\{b_1, b_2, ...b_n\}$. Each vector $b_j$ corresponds to the energy consumption dynamics in a different operation mode of a given device. As shown in figure.3, for example, the energy consumption profiles of the fridge and washer dryer are intermittent and fluctuant even when they are at ON state. Additionally, the energy consumption profiles of a device in different operation models are not linear.

$$\hat{A}_j = \arg\min_{b_i, a_i} \| Y - \hat{B}_{i-1} A \|^2_F + \lambda \sum_{j=1}^{t} f(A) \tag{5}$$

To find the representative basis function, we combine several continuous data segments to serve as the initial dictionary denoted by $\hat{B}_0$

Secondly, we input the initial dictionary $\hat{B}_0$ to estimate the new activations and dictionary until convergence.

$$\hat{B}_j = \arg\min_{b_i, a_i} \| Y - B\hat{A}_{j-1} \|^2_F + \lambda \sum_{j=1}^{t} f(A) \tag{6}$$

subject to $\|a_j\|^2 \leq \delta_0$ and $\|Y - BA\| \leq \sigma$

Here, $\delta_0$ represents the upper bound of ratio 0 in Activation $A$

We also show the detailed process of dictionary training in **Algorihm1**.

**TABLE.1 ALGORIHM OF DICTIONARY TRAINING**

| Algorithm1. Dictionary Training |
|---|
| Input: , $Y_{TrainDate}$, $\hat{B}_0$, $\partial$, $\beta$ |
| Sparse coding pre-training: |
| 1 Initialize $\hat{B}_0$ randomly |
| 2 $\hat{A}_0 \leftarrow \{Y_{TrainDate}, \hat{B}_0\}$ |
| Dictionary training: |
| 3. Iterate until convergence |
| a) $\hat{A}_i \leftarrow \arg\min \| Y - \hat{B}_{i-1}A \|^2_F + \lambda \sum_{i=1}^{t} f(A)$ |
| b) $\hat{B}_j \leftarrow \arg\min \| Y - B\hat{A}_{j-1} \|^2_F + \lambda \sum_{i=1}^{t} f(A)$ |
| subject to $\|a_j\|^2 \leq \delta_0$ and $\|Y - BA\| \leq \sigma$ |
| Output: $B$ |

## E. Randomized Response

For each energy consumption data batch $Y$, sparse coding is used to optimized Dictionary $B$ through training. Then we can get a activations $A = \{a_1, a_2, ... a_n\}$,

Let

$$a_k' = \begin{cases} a_1, & \text{with probability } \frac{c_1}{i} f \\ a_2, & \text{with probability } \frac{c_2}{i} f \\ ...... \\ a_i, & \text{with probability } \frac{c_j}{i} f \\ a_k, & \text{with probability } 1 - f \end{cases}$$

where $f$ is a user-tunable parameter controlling the level of longitudinal privacy guarantee that related to $\varepsilon$.

## F. Data Re-aggregation

Based on the obfuscated activation and dictionary, we can generate the obfuscated energy consumption data $Y'$ as follows:

$Y' = BA'$

Since the sparse coding is a linear mapping, the final obfuscated data $Y'$ satisfies $\varepsilon$-differential privacy (proof is in the next section). However, the final obfuscated result may have a bad impact on the data-utility.

We also show the whole process of the algorithm in **Algorihm2**.

### TABLE.2 RANDOMIZED RESPONSE ALGORITHM

| Algorithm2. Randomized Response Algorithm |
|---|
| Input: , $Y_{TrainDate}$, $Y$, $Y_{TestDate}$, , , $\partial$ , $\beta$ |
| Dictionary training (utility enterprise or fog): |
|   Step 1 Initialize $\hat{B}_0$ and $\hat{A}_0$ |
|   Step 2 Dictionary training: $B \leftarrow training$ |
| Sparse coding(fog or smart meter): |
|   Step 3 computing $A$ according to $B$ |
| $$A \leftarrow B$$ |
| Randomized Response (fog or smart meter): |
|   Step 4 computing $A'$ according to $A$ |
| $$A' \leftarrow A$$ |
| Data Re-aggregation (fog or smart meter): |
|   Step 5 re-aggregation |
| $$Y' = BA'$$ |
| Output: $Y'$ |

## IV. DIFFERENTIAL PRIVACY ANALYSIS

### A. Privacy of Sparse coding

**Theorem 1: Sparse coding still satisfies $\varepsilon$ -differential privacy.**

**Proof:**

Let $Y = y_1, y_2 ..., y_k$ be a batch of energy consumption data. After sparse coding and re-aggregation, $B$ and $A$ are obtained, $Y' = BA$, and $Y' = y'_1, y'_2 ..., y'_k$ is the randomized response generated by the algorithm according to $Y$.

Suppose that $M$ is a privacy machine which satisfies $\varepsilon$ -differential privacy,

$$P_r\{M(Y_1) \subseteq S\} \leq e^{\varepsilon} \times P_r\{M(Y_2) \subseteq S\}$$

The sparse coding algorithm can be viewed as a stable transformation $T$.

$$T: Y' = B_{t \times n} A_{n \times 1}$$

and $\|Y - BA\| \leq \delta_2$, $\delta_2$ is an infinitesimal small number close to 0. We suppose that $Y' = BA$.

So

$$P_r\{M(Y'_1) \subseteq S\} \leq e^{\varepsilon} \times P_r\{M(Y'_2) \subseteq S\}$$

After sparse coding, privacy machine $M$ still satisfies $\varepsilon$ -differential privacy.

**Proof:**

Our scheme $T$ can be expressed as follows:

$$T_{Y_1}: Y1 \xrightarrow{SparseCodeing} Y1'$$

$$T_{Y_2}: Y2 \xrightarrow{SparseCodeing} Y2'$$

Supposed that $T(Y_1) = B_1 A_1 = Y_1$

$$T(Y_2) = B_2 A_2 = Y_2$$

$$|T(Y_1) + T(Y_2)| \leq c|Y_1 + Y_2|$$

In which the sparse coding $T$ is c-stable and the value of c is 1.

According to the property2 of differential privacy, transformation $T_a$ satisfies $\varepsilon$ -differential privacy.

### B. Privacy of Random Response Algorithm

**Theorem 2: The Random Response Algorithm satisfies $\varepsilon$ -differential privacy $\varepsilon = \ln(\frac{1-f}{\delta_0 f})$.**

**Proof:**

$$\text{Let } y_k' = \begin{cases} y'_1, & \text{with probability } \frac{1}{i}f \\ y'_2, & \text{with probability } \frac{1}{i}f \\ ...... \\ y'_i, & \text{with probability } \frac{1}{i}f \\ y'_k, & \text{with probability } 1-f \end{cases}$$

$RR$ is the bounded by $e^{\varepsilon}$

$$RR = \frac{P(y' \in S | y = y_1)}{P(y' \in S | y = y_2)}$$

$$RR \leq max \frac{P(y' \in S | y = y_1)}{P(y' \in S | y = y_2)}$$

considering the sparsity of activations $A$, most $y' = 0$

when $y' = 0$,

$$RR = \frac{P(y' \in S | y = y_1)}{P(y' \in S | y = y_2)}$$

$$RR = (1-f)^2 \times (\delta_0 f)^{-2}$$

$$RR = (\frac{\delta_0 f}{1-f})$$

$$\varepsilon = \ln(\frac{1-f}{\delta_0 f})$$

## V. DIFFERENTIAL PRIVACY ANALYSIS

In this section, using the REDD Data Set [46] and the Nilmtk toolkit [47], we measure the performance of our scheme in terms of the level of privacy preservation.

F1-score is an efficient metric to measure the privacy-preserving level, which can be calculated as follows:

$$F1-score = \frac{2 \times \Pr ecision \times \text{Re} call}{\Pr ecision + \text{Recall}}$$

Here, Precision and Recall represent the positive predictive value and the recall sensitivity respectively. When the F1-score goes high, the application usage patterns can be tracked more accurately.

The commonality shares by Barbosa's scheme, Sankar's scheme and our scheme is the obfuscation of energy consumption data via noise addition. However, these schemes preserve consumer's privacy at different levels. Barbosa adds the noise into energy consumption data directly. Sankar and us add noise into appliance consumption signatures.

As shown in Table.3, we compare the F1-scores of our scheme with Barbosa, Sankar and the classical NILM algorithm (FHMM). From the experiment results, we demonstrate that our scheme's F1-score is lower than the other schemes, which means that it is more difficult to extract the customer's behavior privacy from the noisy energy consumption data in our scheme. Therefore, compared to the other schemes, our scheme has a better privacy-preserving level.

TABLE.3 COMPARISON OF OUR ALGORITHM WITH OTHER TRADITIONAL ALGORITHMS

|  | Our Algorithm | Sankar's[45] Algorithm | Barbosa's[36] Algorithm | without noise |
|---|---|---|---|---|
| Fridge | 0.53 | 0.64 | 0.65 | 0.65 |
| Light1 | 0.58 | 0.65 | 0.64 | 0.71 |
| washer1dryer | 0.44 | 0.50 | 0.49 | 0.54 |
| Light2 | 0.19 | 0.23 | 0.23 | 0.23 |

As shown in Table3, we add noise into different schemes to generate the obfuscated data and then calculate the F1-scores of fridge, light, oven and washer dryer using different with FHMM algorithm. The blue bar represents the F1-score of our scheme, which is higher than the other schemes. The lower F1-scoreamounts to lower precision rate, which means the behavior signatures are harder to extract in our scheme. Thus, by adding noise into the energy consumption signature, our scheme has a better privacy-preserving level.

## VI. CONCLUSION

In this paper, we proposed a Randomized Response Algorithm using the sparse coding to replace Boom filter. The algorithm can achieve differential privacy on batches of data. The algorithm is able to better fulfill the trade-off between privacy and utility in the fog environment against IoT privacy. Therefore, the appliance consumption patterns can be masked even if the adversary has obtained the near real-time load profile. At last, we analyzed the feasibility of our scheme and compared it with other traditional algorithms in IoT. Our algorithm can also be applied to other fields. In future, we will focus on extending the sparse coding to further preserve user's privacy without compromising the data-utility.